\documentclass[%
 reprint,
  superscriptaddress,
 amsmath,amssymb,amsfonts,
 aps,
prl,
 floatfix,
 longbibliography
]{revtex4-2}
\usepackage[utf8]{inputenc}
\usepackage[pdftex]{graphicx} \graphicspath{{}}
\usepackage{float} \usepackage{color}
\usepackage[pdftex,colorlinks=true]{hyperref}
\hypersetup{
  colorlinks=true,
  linkcolor=blue,
  citecolor = blue,
  urlcolor=blue,
}

\usepackage{mathtools}

\DeclarePairedDelimiterX\braket[2]{\langle}{\rangle}{#1 \delimsize\vert #2}
\DeclarePairedDelimiterX\expval[3]{\langle}{\rangle}{#1 \delimsize\vert #2  \delimsize\vert #3}
\DeclarePairedDelimiterX\proj[2]{\delimsize\vert#1\rangle}{\langle#2\delimsize\vert}{ }
\usepackage{siunitx}

\usepackage{tikz}
\definecolor{lime}{HTML}{A6CE39}
\DeclareRobustCommand{\orcidicon}{%
    \begin{tikzpicture}
    \draw[lime, fill=lime] (0,0) 
    circle [radius=0.13] 
    node[white] {{\fontfamily{qag}\selectfont\tiny ID}};
    \draw[white, fill=white] (-0.0625,0.095) 
    circle [radius=0.007];
    \end{tikzpicture}
    \hspace{-2mm}}

\newcommand{\orcidGD}{\href{https://orcid.org/0000-0002-7032-2465}{\orcidicon}}
\newcommand{\orcidLS}{\href{https://orcid.org/0000-0001-7652-9574}{\orcidicon}}

\begin{document}


\title{Emergent Dimensionality in Diluted Power-Law Quantum Walks}
\author{G. A. Dom\'inguez-Castro\orcidGD}
\affiliation{Centro de Nanociencias y Nanotecnología, Universidad Nacional Autónoma de México, Apartado Postal 14, 22800 Ensenada, Baja California, Mexico}
\author{L. Santos\orcidLS}
\affiliation{Institut f\"ur Theoretische Physik, Leibniz Universit\"at Hannover, Appelstr. 2, D-30167 Hannover, Germany}
\date{\today}

\begin{abstract}
We investigate single-excitation quantum walks across the one-to-two-dimensional crossover in diluted power-law hopping models. Using arrays of \(\ell\) coupled chains of length \(L\), we identify the finite-size localization crossover from the longitudinal spreading of an initially localized excitation. The crossover filling is well described by an exponential approach to its two-dimensional value with increasing \(\ell\), defining a characteristic transverse scale \(\ell_e\), and effectively two-dimensional behavior emerges already for \(\ell\ll L\). An independent spectral analysis shows that \(\ell_e\) is set by the transverse extent of the eigenstates participating in the dynamics, providing a microscopic interpretation of the dimensional crossover. Our results reveal an intrinsic eigenstate scale governing emergent dimensionality in diluted systems with power-law hopping, a regime directly relevant to recently developed arrays of polar molecules, Rydberg atoms, and other quantum simulators with long-range interactions.
\end{abstract}

\maketitle


Dimensionality plays a central role in Anderson localization~\cite{PhysRev.109.1492}. For conventional short-range hopping, arbitrarily weak disorder localizes all single-particle states in one and two dimensions~\cite{PhysRevB.22.3519,PhysRevLett.42.673}, whereas in three dimensions localization occurs only beyond a finite disorder strength~\cite{RevModPhys.80.1355}. Two dimensions are marginal: although all states remain localized in the thermodynamic limit, at weak disorder the localization length can greatly exceed experimentally accessible system sizes~\cite{RevModPhys.57.287,PhysRevLett.95.250403}, yielding effectively extended behavior. Increasing disorder reduces the localization length below the system size, giving rise to an effective finite-size localization crossover~\cite{xu2015quantum,PhysRevLett.117.020401}. Localization has been extensively explored at fixed dimensionality~\cite{segev2013anderson,billy2008direct,roati2008anderson,white2020observation}, while finite-width scaling has also been studied in short-range disordered systems, for example in graphene ribbons~\cite{Fan2014}. Much less is known about how localization evolves continuously between different spatial dimensions when the hopping itself is long ranged.

The interplay between dimensionality, hopping range, and effective connectivity becomes particularly important for power-law hopping, \(J_{ij}\propto r_{ij}^{-\alpha}\), where \(r_{ij}\) is the distance between sites and \(\alpha\) controls the hopping range. Power-law disordered systems display localization properties markedly different from those of short-range models, including algebraic localization, multifractality, and anomalous transport~\cite{PhysRevB.106.134208,PhysRevLett.120.110602,PhysRevLett.123.025301,PhysRevB.103.075124,PhysRevB.100.174201,PhysRevA.104.033306}. Random site dilution provides a natural source of correlated off-diagonal disorder: each vacancy removes all hopping processes connected to that site, while power-law couplings retain direct pathways across empty regions. Consequently, the hopping range itself becomes a control parameter for transverse connectivity. How the localization crossover from one to two dimensions develops in such diluted power-law systems, how it depends on the hopping range, and what transverse scale controls it remain largely unexplored.

These ingredients are becoming directly accessible in several quantum-simulation platforms. Dipolar exchange in polar-molecule and Rydberg systems naturally realizes \(1/r^3\) excitation hopping~\cite{yan2013observation,li2023tunable,gunter2013observing,browaeys2020many}, whereas trapped-ion simulators provide tunable power-law interactions~\cite{richerme2014non,jurcevic2014quasiparticle,guo2024site}. In molecular arrays, incomplete filling naturally realizes the dilution considered here~\cite{moses2015creation,mortlock2025multi}, while recent advances in site- and state-resolved microscopy enable the spatial dynamics of individual excitations to be monitored directly~\cite{gross2021quantum,rosenberg2022observation,christakis2023probing,mortlock2025multi}. Together with the geometrical control offered by optical lattices and tweezer arrays, these developments make the dimensional crossover studied here experimentally accessible.



\begin{figure}[t!]
\centering
\includegraphics[width=0.85\columnwidth]{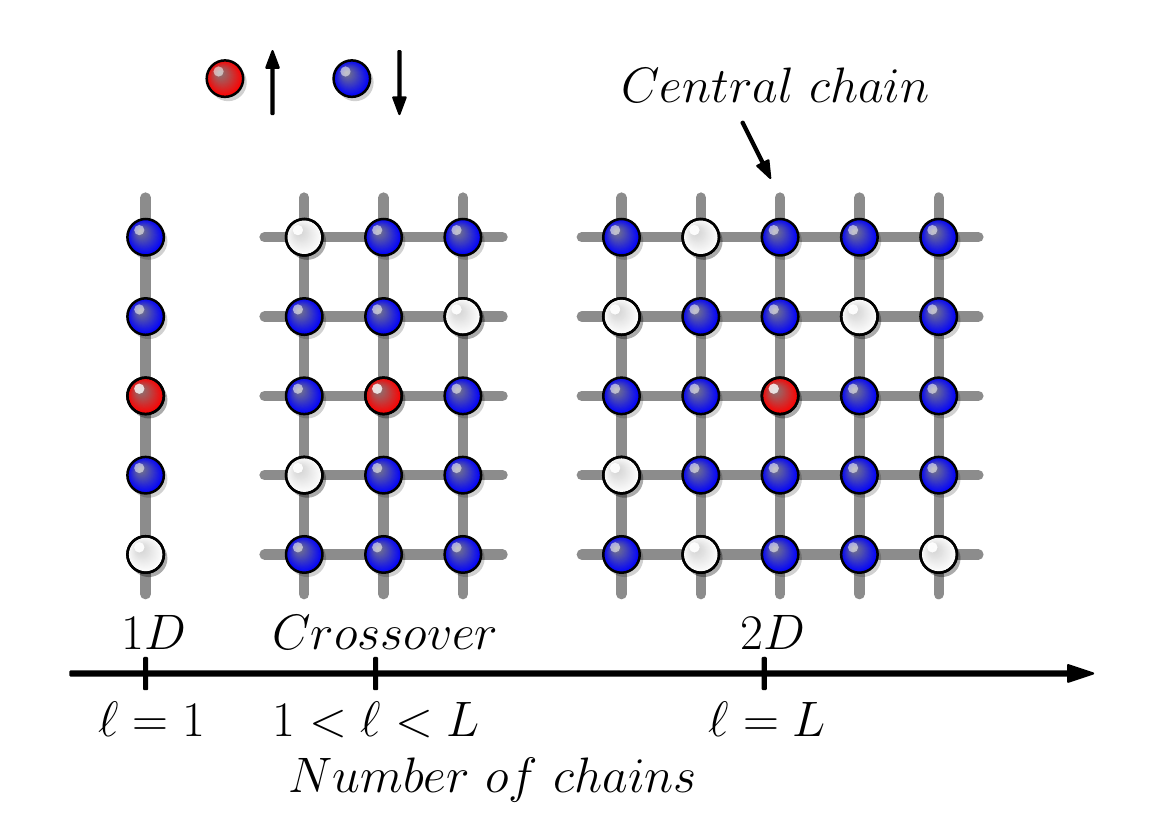}
\caption{Dimensional crossover controlled by the number \(\ell\) of coupled chains. The limits \(\ell=1\) and \(\ell=L\) correspond to a one-dimensional chain and a two-dimensional square lattice, respectively, while \(1<\ell<L\) realizes intermediate quasi-one-dimensional geometries. Red and blue spheres denote particles in the \(\uparrow\) and \(\downarrow\) states, respectively, and white spheres denote vacant sites.}
\label{Fig1}
\end{figure}




\begin{figure*}[t!]
\centering
\includegraphics[width=1.3\columnwidth]{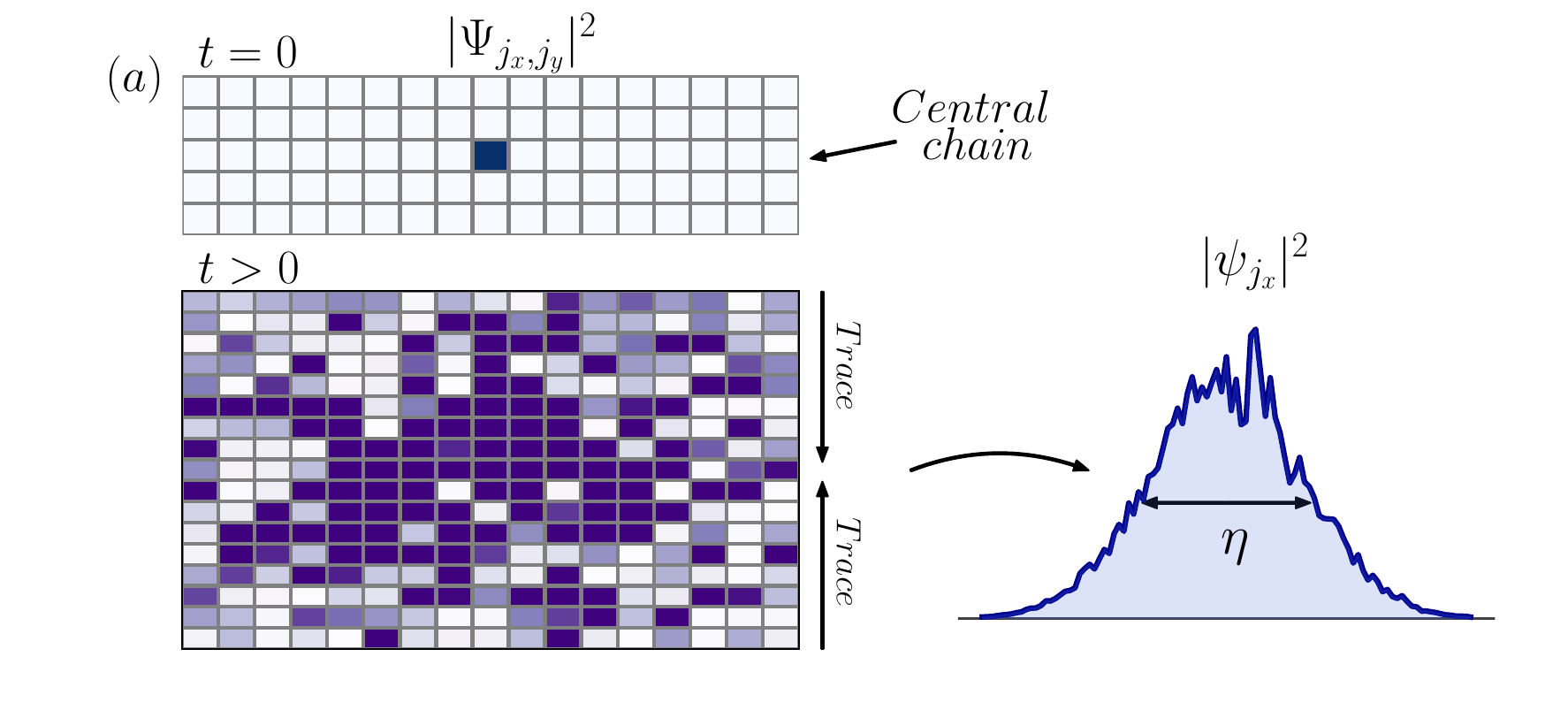}
\includegraphics[width=0.76\columnwidth]{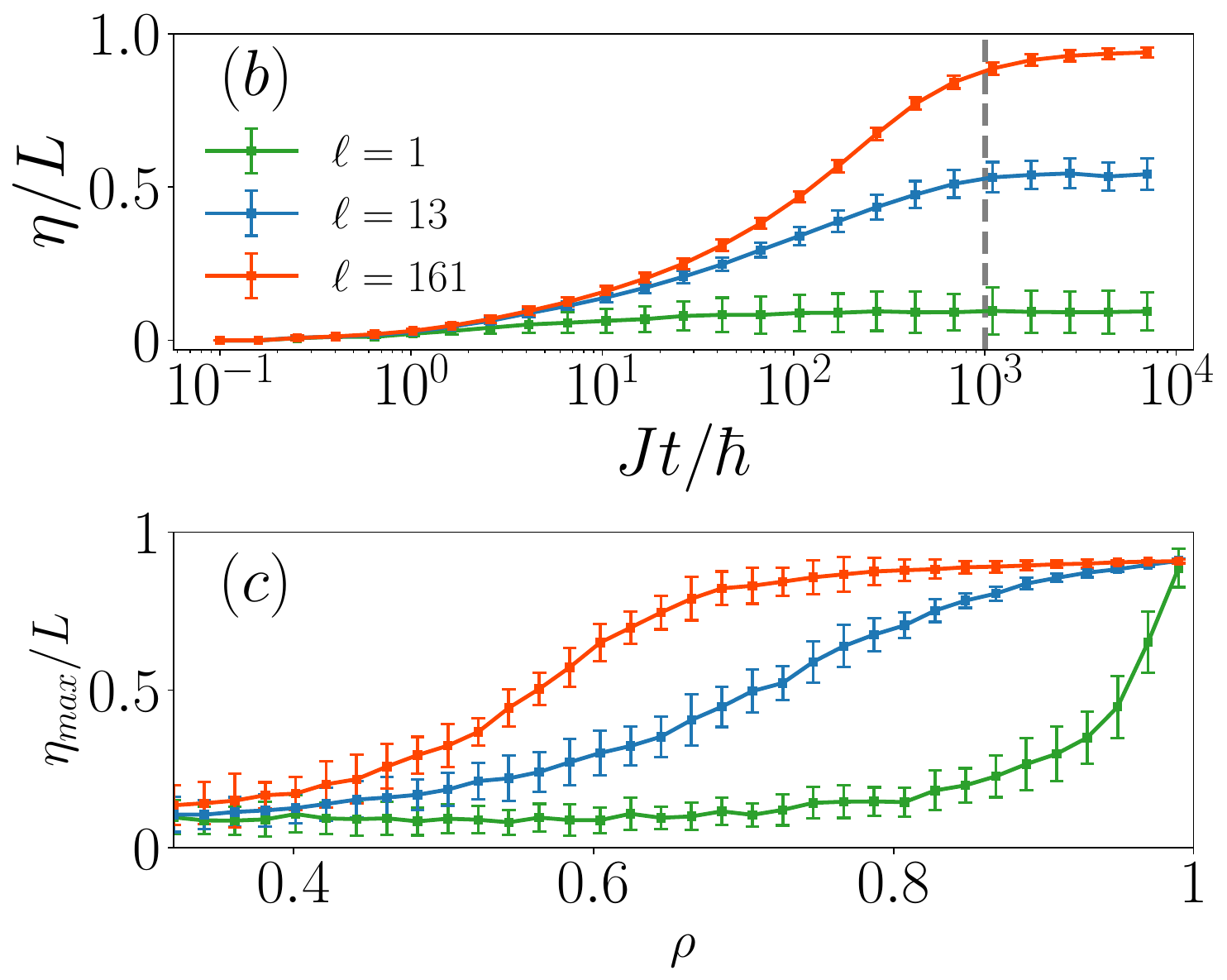}
\caption{(a) Schematic of the reduction to the longitudinal probability distribution. At each time \(t\), the excitation probability \( |\Psi_{j_{x},j_{y}}(t)|^{2}\) is summed over the chain index $j_{y}$, yielding \( |\psi_{j_{x}}(t)|^{2}\). The longitudinal expansion width \(\eta(t)\) is defined as the minimum number of contiguous sites containing \(90\%\) of the excitation probability. (b) Time evolution of \(\eta(t)\) for dipolar hopping (\(\alpha=3\)) at filling \(\rho=0.75\) for three transverse widths \(\ell\). (c) Maximum expansion width \(\eta_{\max}\) versus filling \(\rho\) for the same values of \(\ell\) as in panel (b). Error bars indicate the standard deviation over disorder realizations.
}
\label{Fig2}
\end{figure*}


Here we investigate single-excitation quantum walks on arrays of \(\ell\) coupled chains of length \(L\), interpolating between a one-dimensional chain (\(\ell=1\)) and a two-dimensional square lattice (\(\ell=L\)). From the longitudinal spreading of an initially localized excitation, we extract the finite-size localization crossover as a function of \(\ell\) and \(\alpha\). We find that the crossover filling is well described by an exponential approach to its two-dimensional value with increasing \(\ell\), defining a characteristic transverse scale \(\ell_e\), and that effectively two-dimensional behavior emerges already for \(\ell\ll L\). Moreover, \(\ell_e\) shows a systematic increase as \(\alpha\) decreases. An independent spectral analysis of the two-dimensional system shows that \(\ell_e\) is set by the transverse spatial extent of the eigenstates participating in the dynamics. Thus, the emergence of effective dimensionality in diluted power-law quantum walks is governed not by the total system width, but by an intrinsic transverse eigenstate scale whose magnitude is controlled by the hopping range.



\paragraph{Model.--} We consider pinned two-level particles, \(\{\uparrow,\downarrow\}\), confined to an array of \(\ell\) parallel chains of length \(L\). The lattice spacing along the chains and the interchain separation are both set to unity, \(a=1\). We restrict ourselves to the single-excitation sector, with all occupied sites in the \(\downarrow\) state except for a single \(\uparrow\) excitation. Its dynamics is governed by
\begin{equation}
\hat H=-J\sum_{\mathbf r\neq\mathbf r'}
\frac{\eta_{\mathbf r}\eta_{\mathbf r'}}
{|\mathbf r-\mathbf r'|^\alpha}
|\mathbf r'\rangle\langle\mathbf r|,
\label{Eq1}
\end{equation}
where \(|\mathbf r\rangle\equiv|j_x,j_y\rangle\) denotes an excitation localized at longitudinal site \(j_x\) of chain \(j_y\). The occupation variables \(\eta_{\mathbf r}\in\{0,1\}\) are chosen independently with \(P(\eta_{\mathbf r}=1)=\rho\), where \(\rho\) is the lattice filling~\footnote{We condition the disorder realizations on occupation of the initially excited central site.}.

Equation~\eqref{Eq1} realizes dilution-induced off-diagonal disorder: the hopping between occupied sites retains the deterministic form \(J/|\mathbf r-\mathbf r'|^\alpha\), while a vacancy removes all hopping processes connected to that site. This contrasts with the conventional Anderson model, where disorder is typically introduced through random on-site energies~\cite{PhysRev.109.1492}. Nevertheless, interference among the available hopping paths can suppress transport and induce localization~\cite{PhysRevLett.117.020401,xu2015quantum}. In polar-molecule arrays, this type of disorder arises naturally from incomplete lattice filling~\cite{moses2015creation,mortlock2025multi}.

The transverse width \(\ell\) controls the dimensional crossover: \(\ell=1\) corresponds to a one-dimensional chain, while \(\ell=L\) yields a two-dimensional square lattice [Fig.~\ref{Fig1}]. We use \(L=161\) and odd \(\ell\), ensuring a unique central chain and inversion symmetry about it; similar results are obtained for \(L=101\). The power-law exponent \(\alpha\) depends on the experimental platform: dipolar exchange in polar-molecule and Rydberg systems realizes \(\alpha=3\), whereas trapped-ion simulators allow the interaction range to be tuned. We therefore consider \(2\leq\alpha\leq6\) to explore how the hopping range affects localization across the dimensional crossover.



\begin{figure*}[t!]
\centering
\includegraphics[width=1.0\textwidth]{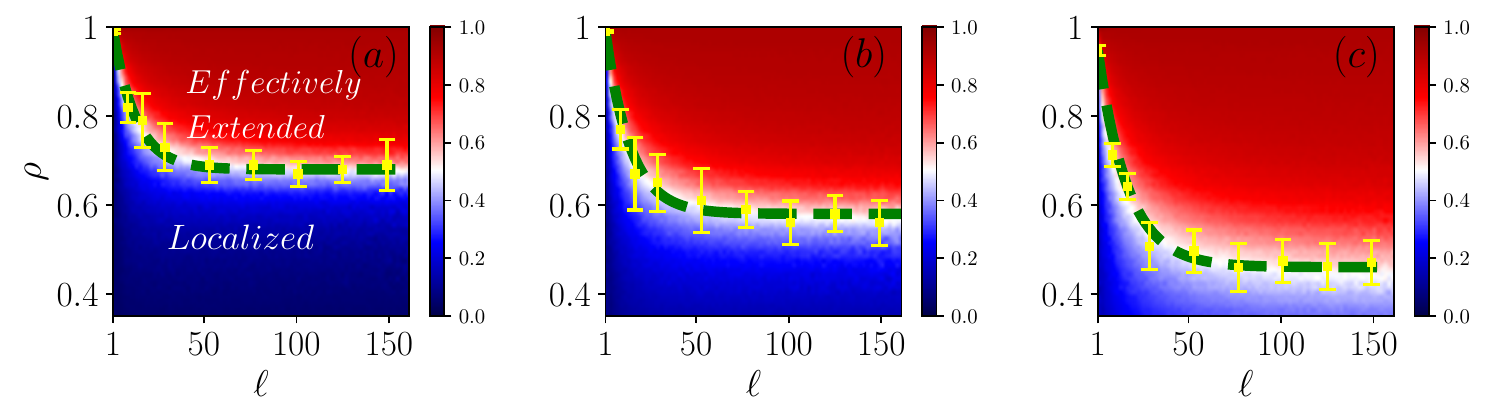}
\caption{(a)--(c) Maximum longitudinal expansion \(\eta_{\max}/L\) versus transverse width \(\ell\) and filling \(\rho\) for \(\alpha=6,3,\) and \(2\), respectively. Yellow markers denote the finite-size localization crossover filling \(\rho_c\), extracted from the inflection-point criterion. Green dashed lines show the exponential fit of Eq.~\eqref{Eq3}. Error bars indicate the standard deviation over disorder realizations.}
\label{Fig3}
\end{figure*}




\paragraph{Quantum walk dynamics.--} We initialize a single \(\uparrow\) excitation at the center of the central chain and obtain its time evolution by exact diagonalization of Eq.~\eqref{Eq1}\footnote{Writing the eigenstates and eigenenergies of Eq.~\eqref{Eq1} as \(|\beta\rangle\) and \(E_\beta\), respectively, we evaluate \(|\Psi(t)\rangle=\sum_\beta e^{-iE_\beta t/\hbar}|\beta\rangle\langle\beta|\Psi(0)\rangle\).}.
Unless stated otherwise, observables are averaged over \(50\)--\(1000\) disorder realizations, depending on system size, with open boundary conditions. To compare quantum walks in arrays with different transverse widths \(\ell\), we trace the two-dimensional excitation probability $|\Psi_{j_x,j_y}(t)|^2$ over the chain index,
$
|\psi_{j_x}(t)|^2=
\sum_{j_y=1}^{\ell}
|\Psi_{j_x,j_y}(t)|^2
$, 
see Fig.~\ref{Fig2}(a). This procedure yields a normalized longitudinal probability distribution defined over the same length \(L\) for all \(\ell\). Following Ref.~\cite{xu2015quantum}, we characterize the spreading by the longitudinal expansion width \(\eta(t)\), defined as the minimum number of contiguous sites containing \(90\%\) of the excitation probability. Localized dynamics corresponds to \(\eta/L\ll1\), whereas \(\eta/L\simeq1\) signals effectively extended dynamics on the finite lattice.

Figure~\ref{Fig2}(b) shows \(\eta(t)\) for dipolar hopping, \(\alpha=3\), at filling \(\rho=0.75\). For a single chain (\(\ell=1\)), disorder strongly suppresses propagation, whereas in the two-dimensional limit (\(\ell=L\)) the excitation spreads over essentially the entire system. Intermediate widths interpolate continuously between these limits; for example, \(\ell=13\) exhibits substantially broader spreading than the one-dimensional chain while remaining less extended than the square lattice. We evolve up to \(Jt/\hbar=10^{4}\), although the same qualitative behavior is already visible at considerably shorter 
times \(Jt/\hbar\sim10^{3}\).

To characterize the dynamics, we define
\[
\eta_{\max}=
\max_{0\leq t\leq t_{\max}}\eta(t),
\qquad
Jt_{\max}/\hbar=10^{4}.
\]
Figure~\ref{Fig2}(c) shows \(\eta_{\max}\) versus filling \(\rho\) for \(\alpha=3\). In one dimension, \(\eta_{\max}\) decreases rapidly upon dilution, whereas in the two-dimensional limit it remains close to \(L\) over a broad high-filling regime. Upon decreasing \(\rho\), \(\eta_{\max}\) develops an inflection point at \(\rho_c^{2D}\simeq0.58\), followed by a rapid suppression of transport. We identify this inflection point with the finite-size localization crossover filling \(\rho_c\), following Refs.~\cite{xu2015quantum,PhysRevB.106.134208}. For intermediate \(\ell\), \(\rho_c\) lies between the one- and two-dimensional limits, providing the dynamical signature of the dimensional crossover analyzed below.



\begin{figure}[t!]
\centering
\includegraphics[width=1.0\columnwidth]{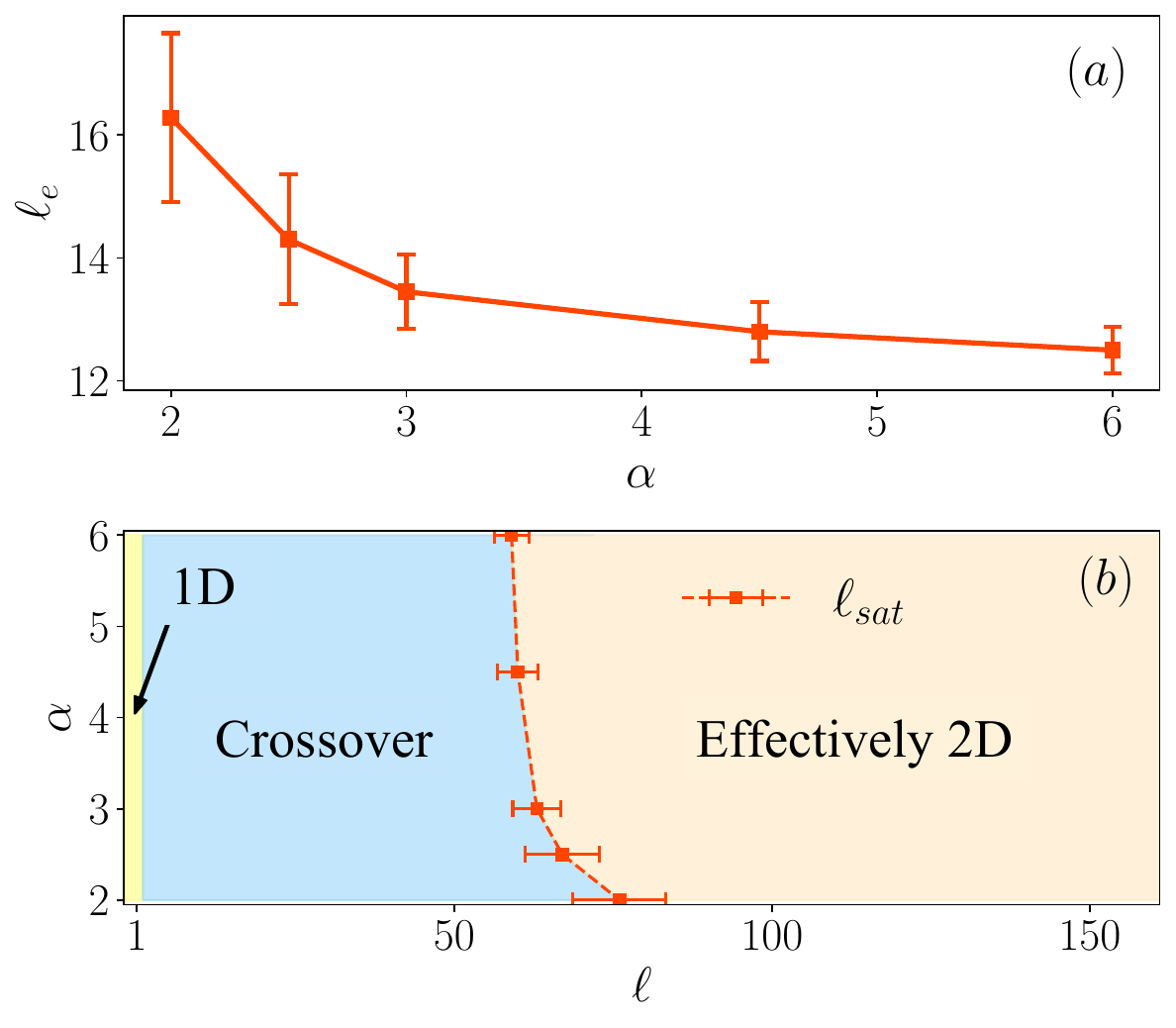}
\caption{(a) Dimensional-crossover scale \(\ell_e\) versus hopping exponent \(\alpha\). (b) 
Operational effective-dimensionality map in the \((\ell,\alpha)\) plane, distinguishing the strictly one-dimensional limit (\(\ell=1\)), the dimensional-crossover regime (\(1<\ell\lesssim\ell_{\rm sat}\)), and the effectively two-dimensional regime (\(\ell>\ell_{\rm sat}\)). Error bars indicate the uncertainty obtained from the fit of \(\ell_e\) and its propagation to \(\ell_{\rm sat}\).}
\label{Fig3-2}
\end{figure}



\paragraph{Characteristic transverse scale.--} Figures~\ref{Fig3}(a)--\ref{Fig3}(c) show \(\eta_{\max}\) as a function of transverse width \(\ell\) and filling \(\rho\) for \(\alpha=6,3,\) and \(2\), respectively. Yellow markers denote the finite-size localization crossover filling \(\rho_c\), extracted from the inflection-point criterion introduced above. For all hopping exponents, increasing \(\ell\) drives a rapid evolution from the strongly localized one-dimensional regime toward the two-dimensional limit, followed by a much weaker dependence on \(\ell\). Thus, effectively two-dimensional localization dynamics emerges already for \(\ell\ll L\).

We quantify this crossover by fitting
\begin{equation}
\rho_c(\ell)=
\left(\rho_c^{1D}-\rho_c^{2D}\right)
e^{-(\ell-1)/\ell_e}
+\rho_c^{2D},
\label{Eq3}
\end{equation}
where \(\rho_c^{1D}\) and \(\rho_c^{2D}\) are determined independently, leaving the characteristic transverse scale \(\ell_e\) as the only fit parameter, see~\cite{SM}. The numerical crossover is well described by Eq.~\eqref{Eq3}. We find \(\ell_e\ll L\), showing that only a finite number of coupled chains is required for the localization dynamics to become effectively two dimensional. Moreover, \(\ell_e\) increases as \(\alpha\) decreases [Fig.~\ref{Fig3-2}(a)], indicating that longer-range hopping requires a larger transverse extent before the two-dimensional limit is reached.

For the shortest-range hopping considered, \(\alpha=6\), we find \(\rho_c^{2D}\simeq0.68\), close to the square-lattice site-percolation threshold \(p_c\simeq0.5927\)~\cite{Stauffer1994,PhysRevE.83.020107,Jacobsen2015}. This proximity reflects the increasing importance of local geometrical connectivity as the hopping becomes shorter ranged. The two thresholds nevertheless remain distinct because power-law hopping preserves long-distance connections, whereas localization is governed by quantum interference~\cite{PhysRevLett.49.486}.

To characterize the transverse width beyond which the dynamics is effectively two dimensional, we introduce an operational saturation width \(\ell_{\rm sat}\), defined by requiring the normalized deviation of the crossover filling from its two-dimensional limit to fall below a tolerance \(\varepsilon\):
\[
\varepsilon=
\frac{\rho_c(\ell)-\rho_c^{2D}}
{\rho_c^{1D}-\rho_c^{2D}},
\qquad
\ell_{\rm sat}=1+\ell_e\ln(1/\varepsilon).
\]
We set \(\varepsilon=10^{-2}\) in Fig.~\ref{Fig3-2}(b), corresponding to a residual deviation of \(1\%\) of the full one-dimensional-to-two-dimensional shift. The resulting diagram separates the strictly one-dimensional limit, the dimensional-crossover regime, and an effectively two-dimensional regime for \(\ell\gtrsim\ell_{\rm sat}\). Since \(\ell_{\rm sat}\) is entirely determined by \(\ell_e\), it serves only as a practical saturation criterion rather than as an independent physical scale.

The increase of the crossover scale with decreasing \(\alpha\) can be understood from hopping processes \(\Delta J(\ell)\) extending beyond the transverse width \(\ell\). For \(\alpha>2\),
\[
\Delta J(\ell)\sim
\int_\ell^\infty dr\,r^{1-\alpha}
\sim \ell^{2-\alpha},
\]
so finite-width corrections decay increasingly slowly as the hopping becomes longer ranged, see~\cite{SM}. The case \(\alpha=2\) is marginal in two dimensions, with \(\Delta J(\ell)\sim\ln(L/\ell)\), reflecting the genuinely long-range character of the coupling and naturally accounting for the particularly broad dimensional crossover observed in this case.



\paragraph{Spectral analysis.--} To uncover the microscopic origin of the dimensional-crossover scale \(\ell_e\), we independently analyze the spatial structure of the single-excitation eigenstates of an \(L\times L\) lattice at filling \(\rho\). For each eigenstate \(\Phi_{j_x,j_y}^{(n)}\), we evaluate its root-mean-square extent along \(y\),
\begin{equation}
\xi_{n,\perp}=
\sqrt{\langle j_y^2\rangle_n-\langle j_y\rangle_n^2},
\qquad
\langle j_y^m\rangle_n=
\sum_{j_x,j_y}j_y^m
|\Phi_{j_x,j_y}^{(n)}|^2 .
\end{equation}
To select the states relevant to the quantum walk, we weight each eigenstate by its overlap with the initially localized excitation,
\begin{equation}
\bar{\xi}=
\sum_n
|\langle\Phi^{(n)}|\Psi(0)\rangle|^2
\xi_{n,\perp},
\end{equation}
where \(|\Psi(0)\rangle\) is the initial state. Importantly, \(\bar{\xi}\) is obtained entirely from the spectrum of the two-dimensional \(L\times L\) system and does not involve the finite-width dependence of \(\rho_c(\ell)\) used to extract \(\ell_e\). Thus, the spectral and dynamical length scales are determined independently. Results are averaged over \(150\) disorder realizations. By symmetry of the square lattice, the disorder-averaged longitudinal and transverse extents coincide.

\begin{figure}[t!]
\centering
\includegraphics[width=0.99\columnwidth]{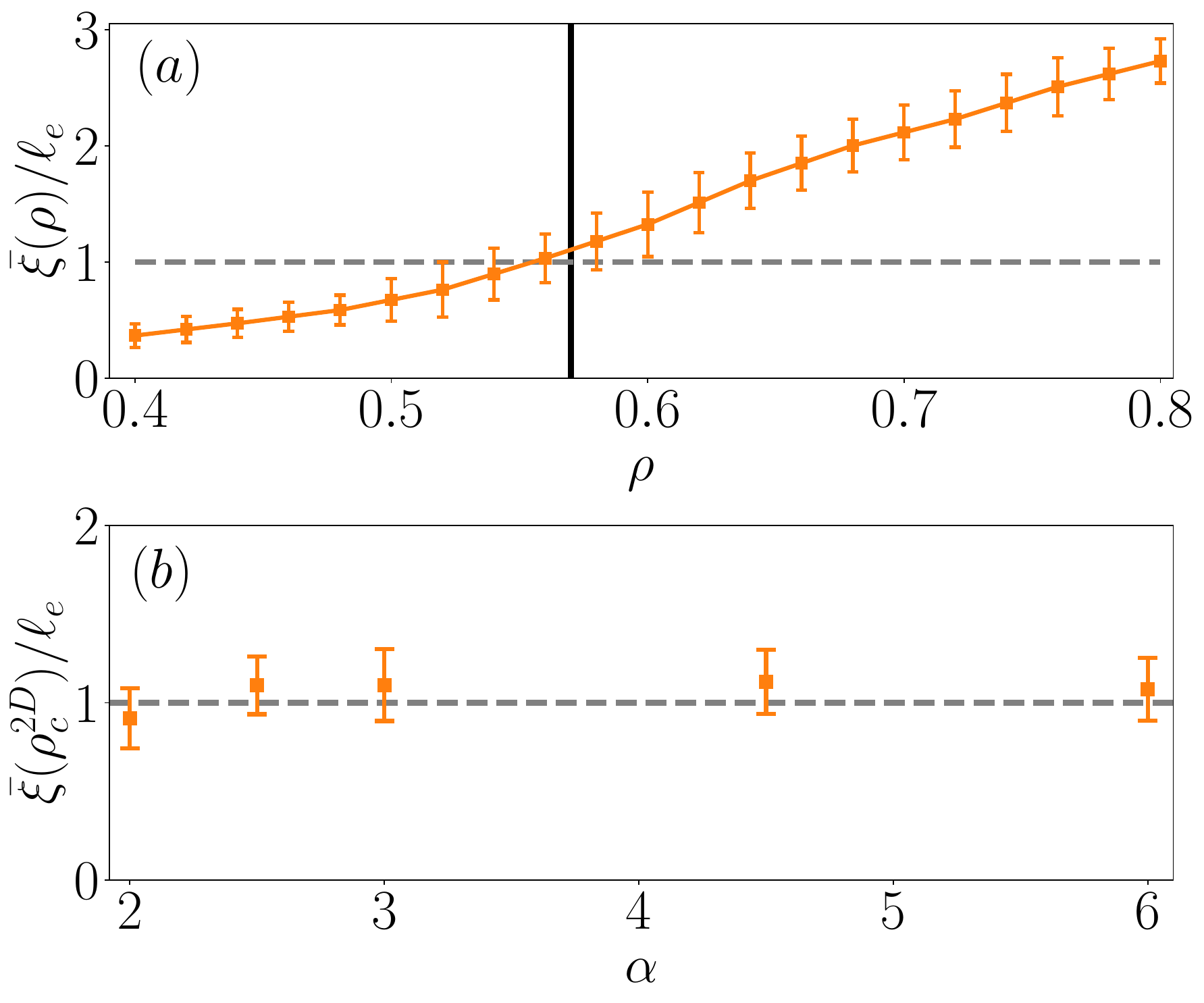}
\caption{(a) Ratio between the eigenstate extent $\bar{\xi}$ and the dimensional crossover scale $\ell_e$, as a function of the lattice filling $\rho$, for $\alpha=3$. (b) Ratio $\bar{\xi}(\rho_c^{\rm 2D})/\ell_e$ for different values of the hopping exponent $\alpha$.  For all $\alpha$, $\ell_e \sim \bar{\xi}(\rho_c^{\rm 2D})$. In both figures the error bars indicate the standard deviation over disorder realizations.}
\label{Fig4}
\end{figure}

Figure~\ref{Fig4}(a) shows the ratio \(\bar{\xi}/\ell_e\) versus \(\rho\) for \(\alpha=3\). The vertical line marks the two-dimensional finite-size localization crossover \(\rho_c^{2D}\), determined independently from the quantum-walk dynamics. Remarkably,
\begin{equation}
\bar{\xi}(\rho_c^{2D})\simeq \ell_e .
\end{equation}
Figure~\ref{Fig4}(b) extends this comparison to \(2\leq\alpha\leq6\). Despite being extracted from distinct calculations, the two scales track each other over the entire range,
\begin{equation}
\ell_e(\alpha)\sim
\bar{\xi}(\rho_c^{2D},\alpha).
\label{Eq6}
\end{equation}
This agreement gives \(\ell_e\), originally introduced through the exponential approach of \(\rho_c(\ell)\) to its two-dimensional limit, a microscopic interpretation: it is set by the transverse extent of the eigenstates participating in the quantum walk. Since \(\ell_e\) grows as \(\alpha\) decreases, the hopping range influences this intrinsic eigenstate scale. Once the array width exceeds it, additional chains only weakly affect the localization crossover. Thus, in diluted power-law systems, the onset of effectively two-dimensional behavior is governed by the spatial structure of the dynamically relevant eigenstates rather than by the total system width.



\paragraph{Conclusions.--} We have studied single-excitation quantum walks across the one-to-two-dimensional crossover in diluted power-law hopping models. The finite-size localization crossover is well described by an exponential approach to its two-dimensional limit with increasing transverse width, defining a characteristic scale \(\ell_e\ll L\). The scale also varies systematically with hopping range, with longer-range hopping requiring larger transverse widths.

An independent spectral analysis shows that \(\ell_e\) is set by the transverse extent of the eigenstates participating in the quantum walk. Together, these results identify a hopping-range-dependent eigenstate scale as the microscopic quantity governing emergent dimensionality in finite, diluted power-law systems. This mechanism is directly relevant to quantum simulators with tunable long-range interactions and site-resolved dynamics.

\acknowledgments
We acknowledge the support of the Deutsche Forschungsgemeinschaft (DFG, German Research Foundation) -- Project-ID 274200144 -- SFB 1227 DQ-mat within the project A04, and under Germany's Excellence Strategy -- EXC-2123 Quantum-Frontiers -- 390837967.

\bibliography{bib-crossover}

\setcounter{equation}{0}
\setcounter{figure}{0}
\renewcommand{\theequation}{S\arabic{equation}}
\renewcommand{\thefigure}{S\arabic{figure}}

\clearpage

\onecolumngrid

\section{Supplemental Material}

\subsection{Exponential fit}

The numerical results of Figs.~\ref{Fig3}(a)--\ref{Fig3}(c) suggest an exponential approach of the crossover filling to its two-dimensional value,
\begin{equation}
\rho_c(\ell)=A e^{-\ell/\ell_e}+C,
\label{Eq1A1}
\end{equation}
where \(\ell_e\) is the characteristic crossover scale. The constants \(A\) and \(C\) are fixed by the boundary conditions
\(\rho_c(1)=\rho_c^{1D}\) and \(\rho_c(L)=\rho_c^{2D}\), yielding
\begin{equation}
\begin{split}
A&=
\frac{(\rho_c^{1D}-\rho_c^{2D})e^{1/\ell_e}}
{1-e^{-(L-1)/\ell_e}}
\simeq
(\rho_c^{1D}-\rho_c^{2D})e^{1/\ell_e},\\
C&=
\frac{\rho_c^{2D}-\rho_c^{1D}e^{-(L-1)/\ell_e}}
{1-e^{-(L-1)/\ell_e}}
\simeq
\rho_c^{2D},
\end{split}
\end{equation}
where the approximations follow from \(L\gg\ell_e\), which is well satisfied for the parameters considered. Equation~\eqref{Eq1A1} therefore reduces to the form used in the main text,
\begin{equation}
\rho_c(\ell)\simeq
(\rho_c^{1D}-\rho_c^{2D})
e^{-(\ell-1)/\ell_e}
+\rho_c^{2D}.
\end{equation}

\subsection{Estimate of long-range hopping contributions}

To understand qualitatively why longer-range hopping requires larger transverse widths to recover effectively two-dimensional dynamics, we estimate the contribution of hopping processes extending beyond a distance \(\ell\). Up to geometry-dependent prefactors, the omitted contribution from a given site can be written as
\begin{equation}
\Delta J(\ell)\sim
\sum_{|\mathbf r|>\ell}
\frac{1}{|\mathbf r|^\alpha}.
\end{equation}
Approximating the two-dimensional lattice by a continuum of radius \(R\sim L\), this becomes
\begin{equation}
\Delta J(\ell)\sim
2\pi\int_{\ell}^{R}dr,r^{1-\alpha},
\end{equation}
and hence
\begin{equation}
\Delta J(\ell)\sim
\begin{cases}
\dfrac{R^{2-\alpha}-\ell^{2-\alpha}}{2-\alpha},
& \alpha<2,\\
\ln(R/\ell), 
& \alpha=2,\\
\dfrac{\ell^{2-\alpha}-R^{2-\alpha}}{\alpha-2},
& \alpha>2.
\end{cases}
\label{EqA_hopping}
\end{equation}

Equation~\eqref{EqA_hopping} highlights the special role of \(\alpha=2\) in two dimensions. For \(\alpha<2\), the contribution of distant sites diverges algebraically in the thermodynamic limit \(R\rightarrow\infty\), whereas for the marginal case \(\alpha=2\) it diverges logarithmically. This divergence is regularized by the finite system size in our calculations, but the contribution of distant hopping processes remains substantial. For \(\alpha>2\), by contrast, \(R^{2-\alpha}\rightarrow0\) as \(R\rightarrow\infty\), giving
\begin{equation}
\Delta J(\ell)\sim\ell^{2-\alpha}.
\end{equation}
Thus, finite-width corrections decay increasingly slowly as \(\alpha\) decreases. This explains why the dimensional crossover extends to larger transverse widths for longer-range hopping and is particularly broad for the marginal case \(\alpha=2\).

\subsection{Additional results}

\begin{figure}[t!]
\centering
\includegraphics[width=0.5\columnwidth]{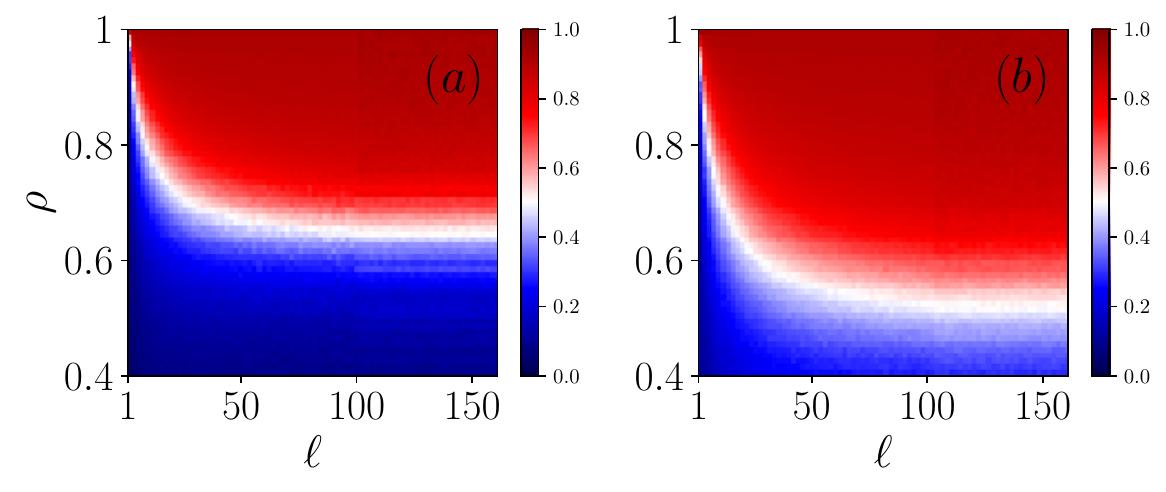}
\caption{Maximum longitudinal expansion \(\eta_{\max}/L\) versus transverse width \(\ell\) and filling \(\rho\) for (a) \(\alpha=4.5\) and (b) \(\alpha=2.5\).}
\label{FigA3}
\end{figure}

To confirm that the behavior discussed in the main text is not restricted to the representative values \(\alpha=6,3,\) and \(2\), Fig.~\ref{FigA3} shows \(\eta_{\max}/L\) as a function of transverse width \(\ell\) and filling \(\rho\) for \(\alpha=4.5\) and \(2.5\). In both cases, the same crossover structure is observed: at small and intermediate \(\ell\), the localization properties depend strongly on the transverse width, whereas for larger \(\ell\) they progressively approach the two-dimensional limit.


\end{document}